\pdfoutput=1
\documentclass[prl,reprint,superscriptaddress,super,superbib,amsmath]{revtex4-1}
\usepackage[separate-uncertainty = true]{siunitx}
\usepackage{graphicx}
\usepackage[export]{adjustbox}

\usepackage{floatrow}
\usepackage[caption = false]{subfig} 
\usepackage{breqn}

\newcommand{\subfigimg}[3][,]{
  \setbox1=\hbox{\includegraphics[#1]{#3}}
  \leavevmode\rlap{\usebox1}
  \rlap{\hspace*{10pt}\raisebox{\dimexpr\ht1-0.5\baselineskip}{#2}}
  \phantom{\usebox1}
}

\makeatletter
\let\cat@comma@active\@empty
\makeatother

\begin{document}

\title{A Hahn-Ramsey Scheme for Dynamical Decoupling and DC Magnetometry with Single Solid-State Qubits}

\author{Nikola Sadzak} \email[Electronic mail: ]{sadzak@physik.hu-berlin.de}
\affiliation{AG Nano-Optik, Institut f\"{u}r Physik and IRIS Adlershof, Humboldt-Universit\"{a}t zu Berlin, Newtonstr. 15, D-12489 Berlin, Germany}

\author{Alexander Carmele}
\affiliation{AG Nichtlineare Optik und Quantenelektronik, Institut f\"{u}r Theoretische Physik, Technische Universit\"{a}t Berlin, Hardenbergstra\ss e 36, D-10623 Berlin, Germany}

\author{Claudia Widmann}
\affiliation{Fraunhofer-Institut f\"{u}r Angewandte Festk\"{o}rperphysik, Tullastra\ss e 72, 79108 Freiburg, Germany}

\author{Christoph Nebel}
\affiliation{Fraunhofer-Institut f\"{u}r Angewandte Festk\"{o}rperphysik, Tullastra\ss e 72, 79108 Freiburg, Germany}

\author{Andreas Knorr}
\affiliation{AG Nichtlineare Optik und Quantenelektronik, Institut f\"{u}r Theoretische Physik, Technische Universit\"{a}t Berlin, Hardenbergstra\ss e 36, D-10623 Berlin, Germany}
\author{Oliver Benson}
\affiliation{AG Nano-Optik, Institut f\"{u}r Physik and IRIS Adlershof, Humboldt-Universit\"{a}t zu
Berlin, Newtonstr. 15, D-12489 Berlin, Germany}

\begin{abstract}Spin systems in solid state materials are promising qubit candidates for quantum information or quantum sensing. A major prerequisite here is the coherence of spin phase oscillations. In this work, we show a control sequence which, by applying RF pulses of variable detuning, allows to increase the spin phase oscillation visibility and to perform DC magnetometry as well. We experimentally demonstrate the scheme on single NV centers in diamond and analytically describe how the NV electron spin phase oscillations behave in the presence of classical noise models. We hereby introduce detuning as the enabling factor that modulates the filter function of the sequence, in order to achieve a visibility of the Ramsey fringes comparable to or longer than the Hahn-echo $T_2$ time and an improved sensitivity to DC magnetic fields in various experimental settings.
\end{abstract}

\maketitle

\begin{figure*}
  \centering
  \begin{tabular}{@{}p{0.4\linewidth}@{\qquad}p{0.25\linewidth}@{}@{\qquad}p{0.25\linewidth}}
    \subfigimg[]{a)}{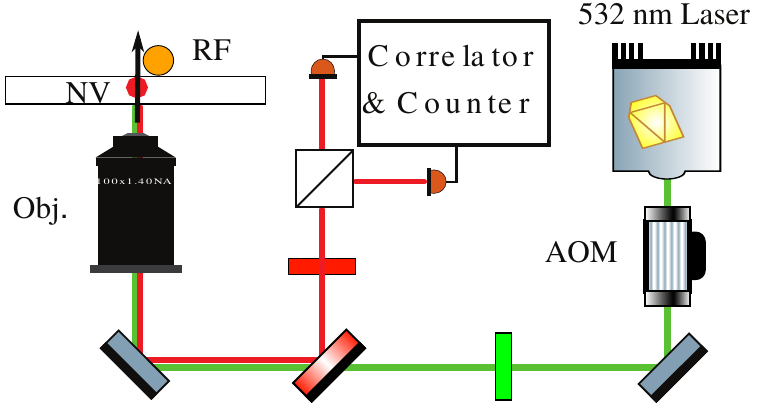} &
    \subfigimg[]{b)}{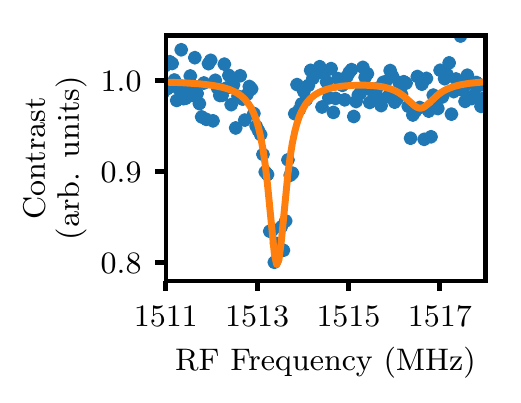} &
    \subfigimg[]{c)}{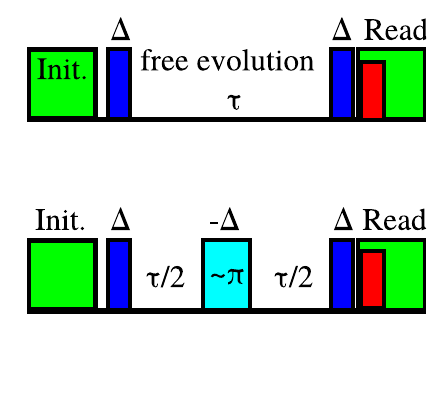}
  \end{tabular}
  \caption{a) Schematic representation of the experimental setup; a [111] diamond plate is illuminated by an AOM-pulsed 532 nm green laser. The emission from the NV centers is collected via an oil immersion objective (NA = 1.40) and sent to a confocal filter and a beamsplitter, terminating on two avalanche photodiodes used for autocorrelation and photon counting measurements. The spins are manipulated by radiofrequency pulses delivered on the diamond by an impedance matched stripline. By applying a static field of $\approx$ 500 G along the NV centers axis, the $^{15}N$ nitrogen nuclear spin of the NV center is hyperpolarized in one of its two states with a polarization ratio $>$ 0.80 (b) as confirmed by the resonance measurement. The scheme in c) shows the Ramsey (top) and Hahn-Ramsey (bottom) sequences, with the green areas representing the polarization and readout laser pulses, and the blue areas representing the microwave pulses controlling the NV centers electron spin, with their detuning indicated as $\pm \Delta$.}\label{fig:basics}
\end{figure*}

 Solid-state qubits are of central importance within the quantum technologies due to their outstanding performance in key fields such as quantum information processing \cite{Stajic2013} and quantum magnetic field sensing \cite{Jones2009,Bal2012}. Several physical systems have been exploited to experimentally realize solid-state qubits such as quantum dots \cite{Hanson2008,Shulman2012,Veldhorst2017, Strauss2019, Carmele2019}, superconductive qubits \cite{Wendin2017,Dicarlo2009}, nuclear spins in materials \cite{Pla2013} and electronic spins in molecules or defect-centers in crystals \cite{Smeltzer2009,Pla2012,GurudevDutt2007,Schlegel2008}. Among the latter, the nitrogen-vacancy (NV) color center in diamond has been extensively investigated due to its exceptional stability and properties observed even at room temperature and in ambient conditions \cite{Kurtsiefer2000}. The NV center has an electron spin triplet that can be optically initialized, readout via fluorescence intensity measurement and controlled with appropriate radiofrequency pulse trains \cite{Jelezko2006,Kennedy2003,Ryan2010}. This has allowed the demonstration of quantum information storage \cite{Maurer2012} and as well to perform magnetic field measurements with high sensitivity and spatial resolution \cite{Degen2017,Sushkov2014,Mamin2012,Loretz2014}. Furthermore, electron spin relaxometry has also been used to probe the behavior of single magnetic domain particles \cite{Gould2014, Sadzak2018} and small ensembles of molecules \cite{Ermakova2013} on the nanoscale. 
 In order to perform sensing in complex physical environments, dynamical decoupling (DD) schemes have been implemented to filter-out the background magnetic noise from the specific target signals. These schemes rely on sequences of precisely timed RF pulses that act as frequency filters and bandwidth selectors. Some basic DD measurements are Ramsey \cite{Ramsey1950} and Hahn-Echo \cite{Hahn1950} schemes, that have been followed by a manifold of other techniques often deriving from the nuclear magnetic resonance (NMR) field \cite{Degen2017}, such as CPMG \cite{Naydenov2011}, XY-n \cite{DeLange2010}, UDD \cite{Uhrig2007,Wang2012}. 
 Most of the currently available decoupling schemes are focused on prolonging the $T_2$ coherence time and are primarly applied to AC magnetometry. Concerning DC magnetometry, the current approaches are  based on using isotopically pure diamonds \cite{Ishikawa2012}, diamond material engineering \cite{Barry2020} or elaborated schemes that circumvent the problem by using spin bath driving, ancillary spins or diamond mechanical rotation \cite{Liu2019, Wood2018, Barry2020}. As these solutions rely on specific experimental configurations, a robust dynamical-decoupling alternative would be particularly interesting for a manifold of applications. 
 Motivated by recent experiments on trapped atoms \cite{Vitanov2015}, we propose here an extension of the Hahn-Ramsey dynamical decoupling scheme, where the detuning of the RF spin control pulses is used to obtain an increased visibility of the electron spin phase oscillations.
 We demonstrate the protocol on single NV centers in bulk diamond, and show a pronounced increase in the spin oscillations coherence time. Furthermore, we give an analytical description of the filter function and of the sequence and provide an estimation of the scheme sensitivity for DC magnetometry, thereby proving that it can be of great importance in a broad range of applications such as quantum sensing, quantum information processing\cite{Nielsen2000} and synchronization \cite{Hodges2013}.
 For our experiment, we use a type [111] CVD-grown delta-doped diamond plate with a ${}^{15}$NV center rich layer. The diamond is placed on a microwave waveguide and the RF pulses are delivered via a 50 \si{\micro\metre} thick copper wire closely located to the surface. From the bottom side, the diamond is accessible via a high numerical aperture (NA = 1.4) oil immersion objective Olympus UPLANSAPO 60X, that is used to optically initialize the nitrogen-vacancies with a 532 nm diode laser source pulsed by an acousto-optic modulator. The same objective collects the fluorescence light that is sent to a confocal setup and a beamsplitter, and finally collimated on two Perkin-Elmer single photon detectors. The experimental setup and some basic measurements are shown in Figure \ref{fig:basics}. 
 
 After identifying a single NV center via autocorrelation measurement and initializing its electron spin in the m${_s = 0}$ state, a typical Ramsey measurement applies two $\pi$/2 pulses separated by a free precession interval, where the electron spin picks up a phase proportional to the external magnetic fields oriented along the NV center quantization axis. The observed signal is the expectation value of the $\sigma_z$ operator $s\left(\tau\right) = Tr\left[\rho\left(\tau\right)\sigma_z\right]$:
 \begin{multline}\label{eq:ramsey}
  s\left(\tau\right) = \langle \uparrow| R^\dagger\left(\theta, \omega_{1} t_p\right) U^{\dagger}\left(0,\tau\right)R^\dagger\left(\theta, \omega_{1} t_p\right)  \sigma_z\\ \times R\left(\theta, \omega_{1} t_p\right)  U\left(0,\tau\right) R\left(\theta, \omega_{1} t_p\right)|\uparrow\rangle,
 \end{multline}
  where $\sigma_z$ is the Pauli spin matrix, U($0,\tau$) is the free evolution operator, $R\left(\theta, \omega_{1} t_p\right)$ the rotation operator for an off-resonant pulse (see supplementary material) and $\theta$ = arctan$\left(\omega_1/\Delta\right)$ the rotation angle \cite{Levitt2000}. Here $\omega_1 = \sqrt{\omega_0^2+ \Delta^2)}$ is the effective precession frequency, with $\Delta$ being the detuning in the driving field, $\omega_0$ the resonant Rabi frequency and $t_p$ the effective pulse duration.

 \begin{figure}[]
    \centering
    \includegraphics[scale = 1.0]{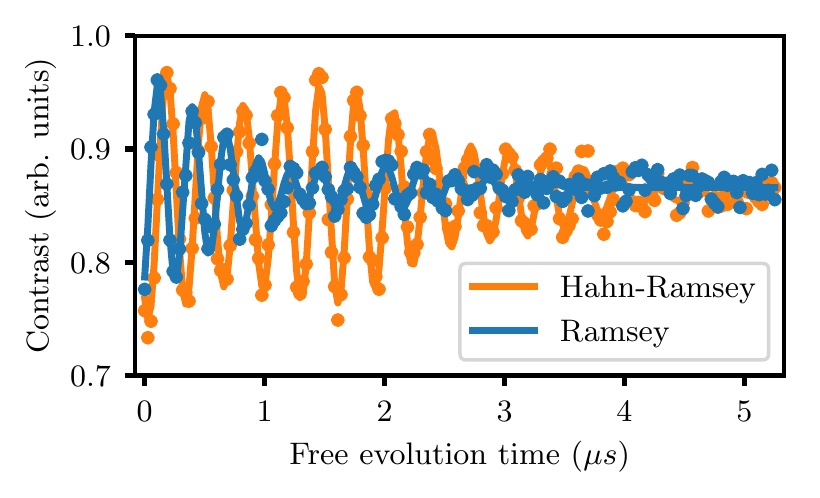}
    \caption{Ramsey (blue) and Hahn-Ramsey (orange) measurements on a single NV center in diamond, with the solid lines showing the fit functions. The beatings are due to a second spin weakly interacting with the NV electron spin. The Hahn-Ramsey fit is done with equation \eqref{eq:hahnramsey} as given in the supplementary material, provided a detuning of $\Delta$ = 4 \si{\mega\hertz} that follows the condition $\Delta < \Omega_{Rabi}$. An exponentially decaying fit of the form $ exp\left[-\left(t/\tau_c\right)^2\right]$ gives a coherence time $T_2^*$ = \SI{1.9 (1)}{\micro\second} for the Ramsey and $T_{2,HR}$ = \SI{3.1 (1)}{\micro\second} for the Hahn-Ramsey.}\label{fig:Single_Hahn_Ramsey}
\end{figure}

 In the case of noisy environments, a time dependent detuning shift arises from the noise contribution to the total magnetic field acting on the NV centers spin. In this situation, the rotating-wave free evolution operator can assume different forms, with the most general one being:
 \begin{equation}\label{eq:noise}
  U_{\Delta}\left(0, \tau\right) = \exp\left[-i\frac{\sigma_z}{2}\int_{0}^{\tau}\left( \Delta + f(t) \right) dt\right],
 \end{equation}
 where f(t) represents the local $\sigma_z$ field at the free precession time t and $\Delta$ is the  detuning term leading to the observation of Ramsey fringes \ref{fig:Single_Hahn_Ramsey}. If we assume f(t) changes on a time scale larger than each acquisition time $\tau$, then a composite pulse sequence may be used to compensate the random phase accumulated for each $\tau$ measurement. 
 This is the principle of the Hahn-echo, where a resonant, refocusing pulse applied at the center of the free evolution time is used to reverse the precession in the second half of the sequence, cancelling the effect of static fields and low-frequency noise but removing spin phase oscillations as well. The Hahn-Ramsey scheme combines the temporal pulse distribution of the Hahn-echo with detuning; the sequence consists of an initial and final $\pi$/2 pulse having a  $\Delta$ detuning, separated by a free precession time $\tau$ and a central pulse of length $\pi$ having instead opposite detuning $-\Delta$ (see Fig. \ref{fig:basics} c). The central, inversely detuned $\pi$ pulse, separates the free precession time in two parts, with the first being described by the operator U$_{\Delta}$($0, \tau$) and the second by U$_{-\Delta}$($\tau, 2\tau$). In this way, the slow contribution of f(t) is cancelled out, leaving nevertheless an oscillating spin phase dependent on $\Delta$ only. The general expression of the Hahn-Ramsey signal is then:
 \begin{multline}\label{eq:hahnramsey}
  s\left(2\tau\right) = \langle \uparrow| R^{\dagger}\left(\theta, \pi/2\right) U^{\dagger}\left(0,\tau\right) R^{\dagger}\left(-\theta, \pi\right)U^{\dagger}\left(\tau,2\tau\right) \\ \times R^{\dagger}\left(\theta, \pi/2\right) \sigma_z R\left(\theta, \pi/2\right)\\ \times U\left(\tau, 2\tau\right) R\left(-\theta, \pi\right) U\left(0,\tau\right) R\left(\theta, \pi/2\right)|\uparrow\rangle.
 \end{multline}
 In order to demonstrate the scheme, we carry out several acquisitions of the Ramsey and Hahn-Ramsey signals on different NV centers in the diamond plate. With the application of a $\approx$ 500 G static field parallel to the defects quantization axis, we achieve a nuclear spin hyperpolarization \cite{Jacques2009} of \SI{80}{\percent} - \SI{90}{\percent} (see Figure \ref{fig:basics}), that allows us to work with an approximate two-level spin-1/2 system. We perform a proof-of-principle measurement by applying on the NV center a well-known artificial noise and comparing the Ramsey and Hahn-Ramsey envelopes; this is shown in the supplementary material.
 
\begin{figure}[]
  \centering
  \includegraphics[scale = 1.0]{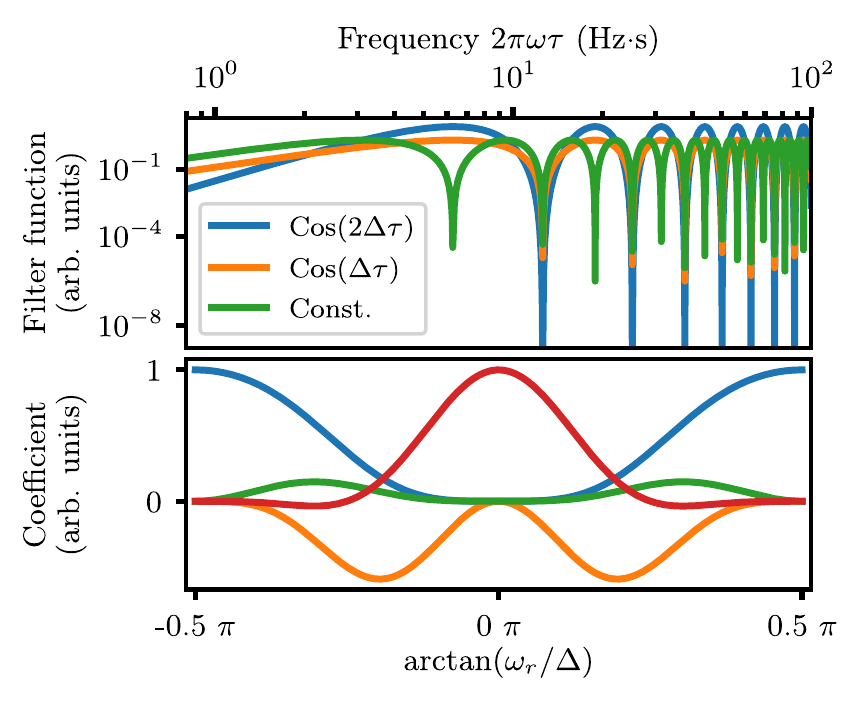}
  \caption{The top figure shows the filter functions contributing to the Hahn-Ramsey signal; the colors represent respectively the non-oscillating part (green), the component oscillating in $\cos\left(\Delta\tau\right)$ (orange) and the component oscillating in $\cos\left(2\Delta\tau\right)$ (blue). The first and the last terms are equivalent to the Ramsey and Hahn-Echo filter functions, while the latter has the same periodicity of the Hahn-echo but different values. The contribution to the overall signal is given by the detuning-dependent weighting coefficients depicted in the bottom figure with their respective colors, with the red line showing the constant term dependent only on the ratio between Rabi frequency and detuning.}\label{fig:simulation}
\end{figure}

\begin{figure}
    \centering
    \includegraphics[scale = 1.0]{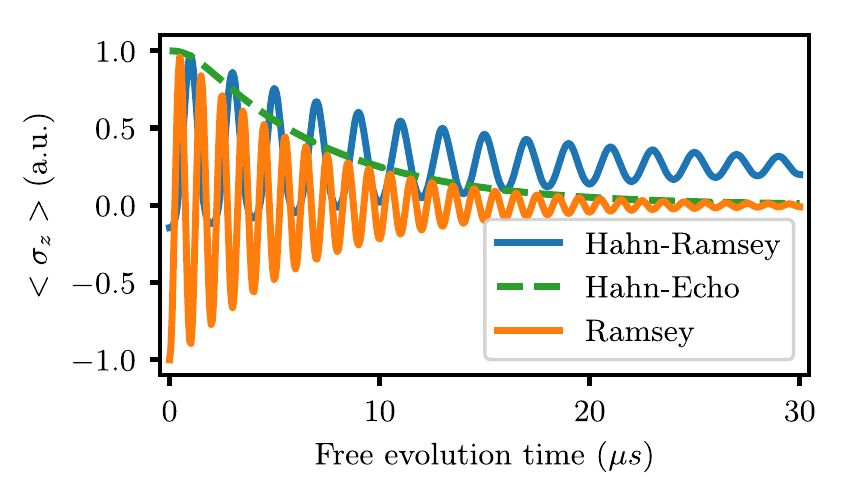}
    \caption{Simulation showing the Ramsey, Hahn-Ramsey and Hahn-echo measurements given the parameters of $\lambda = 2.5$ 1/t, $\Gamma = 2\pi\cdot0.1$ 1/t and $\theta = 0.2 \pi$. For a detuning comparable to the Rabi frequency, the $\cos\left(\Delta\tau\right)$ component contributes significantly to the signal leading to an improved visibility for spin phase oscillations, with a $T_{2,HR}$ time longer than the Ramsey and Hahn-echo times.}\label{fig:comparison_paper}
\end{figure}

 We then proceed to record the Ramsey and Hahn-Ramsey fringes for a specific radiofrequency detuning, as shown in Figure \ref{fig:Single_Hahn_Ramsey}. By testing the sequence effect directly on the NV centers in our diamond, we find that the Hahn-Ramsey is effective in increasing the visibility of the electron spin phase oscillations and the $T_{2,HR}$ time. This is depicted in figure \ref{fig:Single_Hahn_Ramsey}, where an exponential fit indicates an oscillation decay time change from $T_2^*$ \SI{1.9(1)}{\micro\second} to $T_{2,HR}$ \SI{3.1(1)}{\micro\second}. 
 In order to explain the effect of the decoupling sequence, we evaluate Eq.~\eqref{eq:hahnramsey} and derive its form after taking into account the effect of noise processes. 
 We consider f(t) to be a classical magnetic noise represented by a compound Poisson process where the jump times have an exponentially decaying probability density function with a correlation time of 1/$\lambda$ and the jump intensities have a Gaussian distribution with zero average and $\Gamma$ variance. By including this in Eq. \eqref{eq:hahnramsey}, we obtain an expression that links the fringes decoherence profile to detuning and Rabi frequencies, and also to the noise parameters previously defined (see supplementary material).
 After switching to the frequency domain, the expected signal can be expressed as an exponentially decaying function $s\left(\tau\right) = \exp\left[-\chi\left(\tau\right)\right]$\cite{Biercuk2011}, where the exponential argument is the frequency convolution between the control sequence filter function and the spectral density function of the external field or noise process. From Eq.~\eqref{eq:hahnramsey} it can be seen that the opposite detuning for the refocusing pulse with respect to the $\pi/2$ pulses leads to a signal that is the sum of two detuning-induced oscillating components and one non-oscillating component, each one decaying with a different behavior. The overall signal can then be written as: 
\begin{widetext}
\begin{align}\label{eq:detuning}
s_\theta\left(2\tau\right) &= \frac{a^4}{2}\left(1-2b^2\right) +\frac{a^2 b^4}{2} \exp\left[-\frac{2\lambda\Gamma^2}{\pi}\int_{0}^{\infty}  \frac{d\omega/\omega^2}{\omega^2 + \lambda^2}\sin^2\left(\omega\tau\right)  \right]\\ \nonumber
&- 2 a^4 b^2 \cos\left(\Delta\tau\right)\exp\left[-\frac{2\lambda\Gamma^2}{\pi}\int_{0}^{\infty}  \frac{d\omega/\omega^2}{\omega^2 + \lambda^2} \sin^2\left(\frac{\omega\tau}{2}\right)\right]
+ \frac{b^4}{2}\left(a^2+1\right)\cos\left(2\Delta\tau\right) \exp\left[-\frac{8\lambda\Gamma^2}{\pi}\int_{0}^{\infty}  \frac{d\omega/\omega^2}{\omega^2 + \lambda^2}\sin^4\left(\frac{\omega\tau}{2}\right)\right]
\end{align}
\end{widetext}
with $a = \cos(\theta)$ and $b = \sin(\theta)$. We find out that the Hahn-Ramsey signal is the superposition of three decaying components, weightened by detuning related coefficients (see Fig. \ref{fig:simulation}). The term with the spin phase oscillating as $\cos\left(2\Delta\tau\right)$ is associated with a Hahn-Echo type of filter function, that means the coherence decays as in the standard Hahn-echo sequence, while the non-oscillating term shows a Ramsey-like decoherence. The term oscillating instead as $\cos\left(\Delta\tau\right)$ has a different behavior; its filter function has a similar periodicity of the Hahn-Echo but different magnitude, and gives a longer decoherence time. By choosing an opportune detuning one can select how each component is contributing to the total signal. The $\cos\left(\Delta\tau\right)$ component not only provides, in certain cases, a longer spin phase coherence time with respect to the resonant Hahn-echo (see Fig. \ref{fig:comparison_paper}), but it can also be used to measure small DC magnetic fields that act as a bias and shift the detuning of the $\pi$ and $\pi/2$ pulses asymmetrically. In this case the sensitivity, maximized for $\theta_p = 0.2\pi$, is:
\begin{equation}
 \eta_{HR} \propto\frac{1}{3\pi\gamma_e\sqrt{T_{2,HR}}}
\end{equation}

In conclusion, we have described and experimentally demonstrated a Hahn-Ramsey type of dynamical decoupling sequence on NV centers in diamond. By opportunely inverting the detunings in the RF control sequence, we are able to show that the HR scheme is effective in providing a better suppression of low-frequency noise with respect to the Ramsey scheme, approaching the Hahn-echo limit when the detuning is smaller than the Rabi frequency. This can be exploited to obtain an improved spin phase oscillation visibility. When the detuning magnitude is instead comparable to the Rabi frequency, the Hahn-Ramsey can provide even longer Ramsey fringes decay times and may also be used for DC magnetometry, providing a better sensitivity than the standard Ramsey interferometry. 
\\
\\
 A.K. and A.C. gratefully acknowledge support from the Deutsche Forschungsgemeinschaft (DFG) through the project B1 of the SFB 910 and from the European Union’s Horizon 2020 research and innovation program under the SONAR Grant Agreement No. 734690. N.S. gratefully acknowledges funding from the Deutsche Forschungsgemeinschaft (DFG) through the SFB 951.

\bibliography{bibliography_hahn_ramsey.bib}
 
\end{document}


\title{Supplementary Material}
\maketitle

\section{Pulse operator}

We define the general operator for an off-resonant RF pulse \cite{Levitt2000}:
\begin{equation}
    R\left(\theta, \omega_{1} t_p\right) = R_y\left(\theta\right) R_z\left(\omega_{1} t_p \right) R_y\left(-\theta\right).
\end{equation}
Here, $ t_p $ is the pulse duration, $\omega_1$ is the effective precession frequency defined as $\omega_1 = \sqrt{\omega_0^2+ \Delta^2}$ with $\Delta$ being the RF detuning with respect to the spin transition frequency, $\omega_0$ is the resonant Rabi frequency, and $\theta$ is defined as $\theta = \arctan\left(\omega_1/\Delta\right)$.

\section{Hahn-Ramsey sequence: Derivation of the signal}

In the following we show the detailed derivation of the $<S_z>$ expression (main text Eq. 3) for the Hahn-Ramsey sequence given the specific conditions that are mentioned in the main text; the final formula (equivalent to Eq. 4 of the main text) reads:
\begin{align}\label{eq:HR}
   \ew{ S_z (2\tau)} &= \frac{a^4}{2}\left(1-2b^2\right) + \frac{a^2 b^4}{2} e^{-2\left(F_1+\delta F\right)}\notag \\
    &- 2 a^4 b^2 \cos\left(\Delta\tau\right) e^{-F_1} + \frac{b^4}{2}\left(a^2 + 1\right) \cos\left(2\Delta \tau\right)e^{-2\left(F_1-\delta F\right)},
\end{align}
where $ a=\cos\Theta $ and $ b=\sin\Theta $ and $ \Theta $ is the rotation
angle given from the experiment and $ \Delta=\omega_s-\omega_0$ describes the detuning between the driving
microwave field and the spin eigenfrequency.
The noise contribution are given with:
\begin{align}
F_1 :=& 
\frac{1}{2}
\int_0^\tau dt_1 \int_0^\tau dt_2
\avg{F(t_1)F(t_2)} \\
\delta F :=& 
\frac{1}{2}
\int_0^\tau dt_1 \int_\tau^{2\tau} dt_2
\avg{F(t_1)F(t_2)},
\end{align}
for which we use the noise correlation:
\begin{align}
\avg{F(t_1)F(t_2)} =&
\Gamma^2 e^{-\lambda |t_1-t_2|},
\end{align}
and therefore evaluate the dephasing constants to:
\begin{align}
F_1 =& \frac{\Gamma^2}{\lambda^2}
\left[ 
\lambda \tau
+e^{-\lambda\tau}
-1
\right] , \\
\delta F =& \frac{1}{2} \frac{\Gamma^2}{\lambda^2}
\left[ 
1-
2
e^{-\lambda\tau}
+
e^{-2\lambda\tau}
\right].
\end{align}

This process is associated with an ergodic Ornstein-Uhlenbeck 
process:
\begin{align}
F_t =& F_0 e^{-\lambda t} + \Gamma^2\lambda \int_0^t e^{-\lambda(t-s)} dW_s
\end{align}
where $ W $ describes a Wiener process with $ \avg{dW_s}=0 $ and
$ \avg{dW_s dW_t} = \delta(t-s) $.
Due to this process, we can evaluate an exponential of a stochastic variable in a
Gaussian way:
\begin{align}
\avg{\exp\left[\int_0^\tau F(t_1)dt_1 \right]}
= \exp\left[-\frac{1}{2}\int_0^\tau \int_0^\tau \avg{F(t_1)F(t_2)} dt_1 dt_2\right].
\end{align}

Given these parameters, we define now the Hahn-Ramsey sequence via
the canonical spin operators $ [S_i,S_j]=i\epsilon_{ijk}S_k $:
\begin{align}
\ket{\Psi_f} =&
R(\Theta,\pi/2)
e^{-i2gS_z}
R(-\Theta,\pi)
e^{-i2fS_z}
R(\Theta,\pi/2)
\ket{0}
\end{align}
with $ R(\alpha,\beta)=\exp[-iS_y\alpha] \exp[-iS_z\beta]\exp[iS_y\alpha]$
and $ f=(\Delta\tau+\int_0^\tau F(t_1) dt_1)/2 $ and
$ g=(-\Delta\tau+\int_\tau^{2\tau} F(t_1) dt_1)/2 $.
To calculate the signal, we switch into the density matrix picture:
\begin{align}
\ew{ S_z(2\tau)} =&
\text{Tr}
\left[
\rho(2\tau) R^\dg(\Theta,\pi/2)S_zR(\Theta,\pi/2)
\right].
\end{align}
The density matrix can be calculated analytically:
\begin{align}
\rho(2\tau) =&
\frac{1}{2}
\begin{pmatrix}
\substack{a^2+a^4+b^4 \\ -2ab^2[a\cos(2f)+\sin(2f)]} & 
\substack{-a^2b(i+a)e^{-2i(f+g)}-2a^3be^{-i2g} \\
 +b^3(a-i) e^{i2(f-g)} } \\
\substack{-a^2b(-i+a)e^{2i(f+g)}-2a^3be^{i2g} \\
 +b^3(a+i) e^{-i2(f-g)}} &
\substack{b^2+2a^2b^2 \\ +2ab^2[ a \cos(2f)+\sin(2f)]}
\end{pmatrix},
\end{align}
which fulfills $ \text{Tr}[\rho]=1 $ and $ \rho=\rho^\dg $.
With
\begin{align}
R^\dg(\Theta,\pi/2)S_zR(\Theta,\pi/2) =& 
\begin{pmatrix}
a^2 & b(a-i) \\
b(a+i) & -a^2 
\end{pmatrix},
\end{align}
we can compute the expectation value:
\begin{align}
\ew{ S_z(2\tau)} =&
a^2 \left[
\bra{0}\rho(2\tau)\ket{0}
-
\bra{1}\rho(2\tau)\ket{1}
\right] \\
& +2ab \text{Re}
\left[
\bra{0}\rho(2\tau)\ket{1}
\right] 
- 2b 
\text{Im}
\left[
\bra{0}\rho(2\tau)\ket{1}
\right].
\end{align}
Taking the matrix elements, we yield the formula given in the 
main text.
To calculate the signal, the noise correlation is used:
\begin{align}
\avg{ \cos(2f) } =& \cos(\Delta \tau) e^{-F1} \\
\avg{ \sin(2f) } =& \sin(\Delta \tau) e^{-F1} \\
\avg{ \exp[-2i(f+g)] } =& e^{-2(F1+\delta F)} \\
\avg{ \exp[ 2i(f-g)] } =& e^{-2(F1-\delta F)} e^{i\Delta2\tau}  \\
\avg{ \exp[-2i g ] } =& e^{-F1} e^{i\Delta\tau}. 
\end{align}
Therefore, in the long-time limit $ \tau\rightarrow\infty $, the density matrix
is a mixed state without off-diagonal elements:
\begin{align}
\lim_{\tau \to \infty} \rho(\tau) =&
\frac{1}{2}
\begin{pmatrix}
a^2+a^4+b^4  & 0 \\
0 & b^2+2a^2b^2 
\end{pmatrix},
\end{align}
since $ \exp[F_1] \to 0 $.

\section{Hahn-Ramsey with Artificial Noise}

\begin{figure*}[h]
    \subfigimg[]{a)}{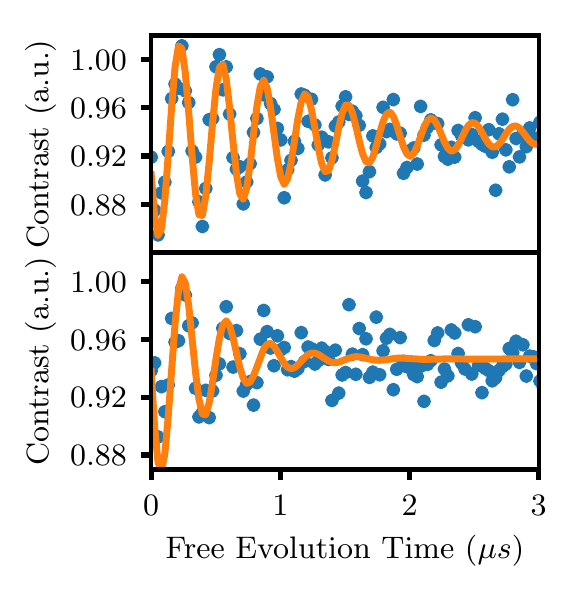}\hspace*{-0.7em} 
    \subfigimg[]{b)}{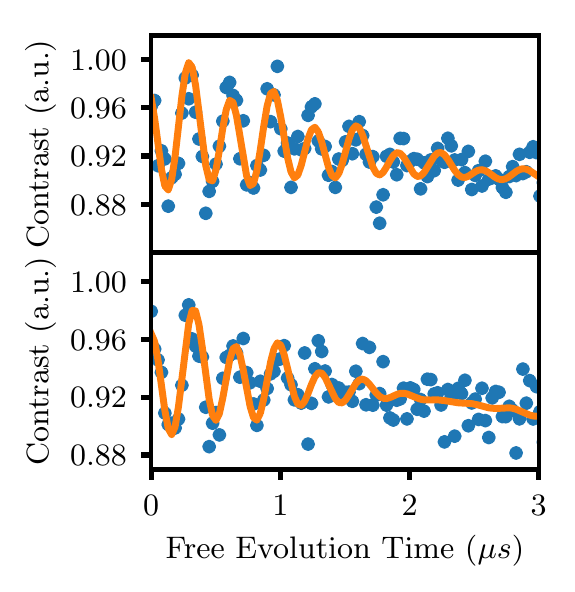}\hspace*{-0.7em} 
    \subfigimg[]{c)}{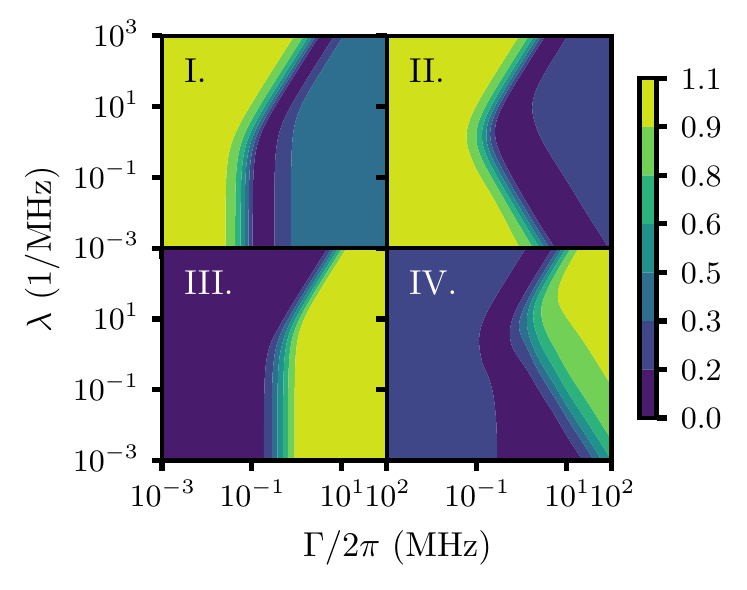}
   \caption{a) Ramsey measurements on a single NV center without (top) and with (bottom) the artificial Poissonian noise. b) Hahn-Ramsey measurement without (top) and with (bottom) the artificial noise. By comparing the two, it is possible to observe that the Hahn-Ramsey sequence provides longer coherence time. In c), the strength and correlation parameters for the intrinsic noise are derived from an optimal-fit method showing the normalized residuals for the chosen parameter space, where a value close to zero indicates a better fit to the experimental results. In I and III, the Ramsey sequences are compared, with the color plot evidencing the optimal parameter area for the bare NV measurement without (I) and with (III) the artificial noise. In II and IV, the same is done with the Hahn-Ramsey sequence. By combining the optimal parameters for Ramsey and Hahn-Ramsey, it is possible to conclude that the intrinsic noise has a correlation time higher than $\approx$ 0.1 \si{\micro\second}, as this is consistent with both of the measurements.}\label{fig:test}
\end{figure*}

In order to test the sequence, we choose an NV center with similar $T_2^*$ and $T_{2,HR}$ coherence times . We therefore select an NV center with a \SI{1.6(1)}{\micro\second} $T_2^*$ time; we generate a magnetic noise pattern by means of a small solenoid having its symmetry axis aligned with the NV center quantization axis. In this way we are able to apply a noise with well defined statistical properties that are independent of the NV center environment. We measure the Ramsey and Hahn-Ramsey decoherence profiles with and without the artificial noise, and use the model described in the paper (Eq. 6-9) to explain the observations (see Figure \ref{fig:test}). We find that the Hahn-Ramsey sequence is effective in partially compensating the induced noise and the Hahn-Ramsey and Ramsey decays are well fitted by the theoretical model. By deconvoluting the environmental component from the induced Poissonian noise, we are able to gain an insight into the strength and dynamics of the environmental fluctuations, and the plots in Figure \ref{fig:test} c) show the resulting optimal fit parameters for the intrinsic noise.

\newpage\null\thispagestyle{empty}\newpage

\section{Derivation of the sensitivity}

We start from the expression of the Hahn-Ramsey signal $s\left(2\tau\right)$ as defined in Eq. \eqref{eq:HR} of the supplementary. Now, this is calculated for a $+\Delta$ detuning for the two $\pi/2$ control pulses and an opposite $-\Delta$ detuning for the refocusing central $\pi$ pulse. By adding a small static magnetic field to the physical system, the spin transition resonance line shifts and the pulse detunings become asymmetric. Specifically the $\pi/2$ pulses in the rotating wave picture become $\Delta-\epsilon$ detuned, while the central refocusing pulse transforms to a $-\Delta+\epsilon$ detuning. By recalculating the signal for these detunings, we get:
\begin{align}
 \ew{S_z\left(2\tau, \epsilon\right)} & = \frac{a^4}{2}\left(1-2b^2\right) + \frac{a^2 b^2}{2}\exp(-2(f_1+\delta f))\left[b^2\cos\left(2\epsilon\tau\right) - 2a\sin\left(2\epsilon\tau\right)\right] \notag \\ 
 &- 2 a^3 b^2\exp\left(-f_1\right)\left[\cos\left(\Delta\tau\right)\left(a\cos\left(\epsilon\tau\right) + \sin\left(\epsilon\tau\right)\right)\right]\notag \\ 
 &+ \frac{b^4}{2}\left(a^2 + 1\right)\cos\left(2\Delta\tau\right)\exp\left(-2(f_1-\delta f)\right).
\end{align}
In order to estimate the sensitivity, we calculate the derivative of the previous equation with respect to $\epsilon$:
\begin{align}
 \frac{\partial\ew{S_z\left(2\tau, \epsilon\right)}}{\partial\epsilon} & = -2a^3b^2\exp\left(-f_1\right)\left[\tau\cos(\Delta\tau)\left(\cos(\epsilon\tau)-a\sin(\epsilon\tau)\right)\right] \notag\\
 &- a^2b^2\exp(-2(f_1+\delta f))\tau\left[2a\cos(2\epsilon\tau) + b^2\sin(2\epsilon\tau)\right].
\end{align}
Following now the approach of reference \cite{MyPham2015}, we consider u and v to be the photon intensity for a phase accumulation in the free precession time of 0 and $\pi$. For the NV center, this is equivalent to 'bright' and 'dark' states. We define then $\alpha = (u-v)/(u+v)$ as the measurement contrast and $\beta = (u+v)/2$. The optical signal collected during a magnetic field measurement is hence given by:
\begin{equation}
 S(2\tau) = \frac{u+v}{2} + \frac{u-v}{2}\ew{s(2\tau)}.
\end{equation}
The smallest detectable field (B) is given by:
\begin{equation}
 \delta B_{min} = \frac{\delta S}{max\left|\frac{\partial S}{\partial B}\right|}
\end{equation}
with $\epsilon = 2\pi\gamma_e B$, $\gamma_e$ the NV electron gyromagnetic ratio, and $\delta S = \sqrt{\beta}$.
Combining these equations, we get that the minimal detectable field for the Hahn-Ramsey is:
\begin{equation}
 \delta B_{min} = \frac{1}{3\pi\gamma_e\tau \alpha \sqrt{\beta}}
\end{equation}
where the sensitivity follows as:
\begin{equation}
 \eta_{HR} \propto \frac{1}{3\pi\gamma_e\alpha\sqrt{T_{2,HR}}}.
\end{equation}
The Hahn-Ramsey for DC magnetometry performes better than the Ramsey especially for the cases of fast environmental noise and low RF pulse powers (Rabi frequencies).

\section{Bloch Sphere Trajectory}

\begin{figure}[h]
 \includegraphics[scale = 1.0]{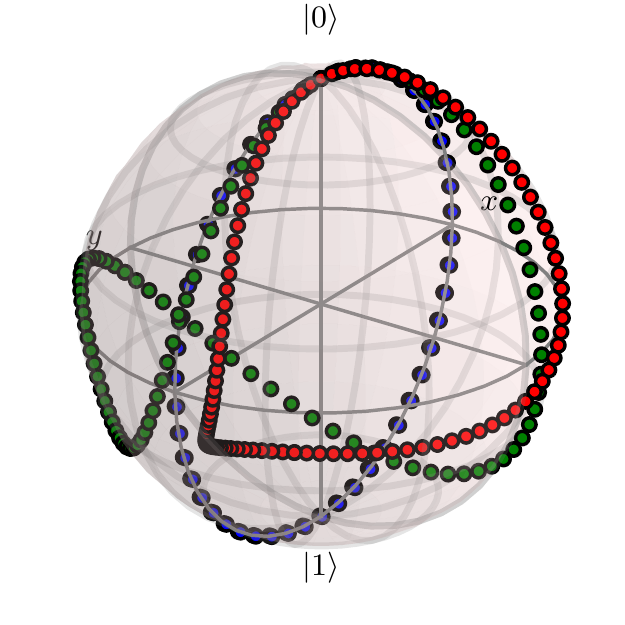}
 \caption{Single fringe, noiseless Bloch sphere trajectories for the Hahn-Ramsey scheme with $\theta \simeq 0.5 \pi$ for the blue dots,  $\theta = 0.3 \pi$ for the green dots, and $\theta = 0.2 \pi$ for the red dots.}\label{fig:bloch}
\end{figure}

Figure \ref{fig:bloch} depicts few trajectories of the spin qubit state under the Hahn-Ramsey scheme and for a single fringe without any overlapping noise process. The signal as calculated in the previous sections is here obtained by projecting the final state on $\sigma_z$. Different colors are used to evidence the different detuning-to-Rabi ratios ($\theta$ angles) and how they contribute to the observed signal shape.  

\bibliographystyle{apsrev4-1}
\bibliography{SupMatBib.bib}